\documentclass[runningheads]{llncs}
\usepackage{graphicx}
\usepackage{hyperref}
\usepackage{amsmath}
\usepackage{dirtytalk}

\begin{document}
\title{Investigating the Origins of Fractality Based on Two Novel Fractal Network Models}

\titlerunning{Investigating the Origins of Fractality Based on Two Novel Network Models}
%
%
\author{Enikő Zakar-Polyák\inst{1}\orcidID{0000-0001-5655-4940} \and
Marcell Nagy\inst{1}\orcidID{0000-0001-5666-7777} \and
Roland Molontay\inst{1,2}\orcidID{0000-0002-0666-5279}}
\authorrunning{E. Zakar-Polyák et al.}
%
\institute{Department of Stochastics, Institute of Mathematics,
Budapest University of Technology and Economics, Műegyetem rkp. 3., H-1111 Budapest, Hungary \and
ELKH-BME Stochastics Research Group, Műegyetem rkp.
3., H-1111 Budapest, Hungary}
\maketitle              
\begin{abstract}
Numerous network models have been investigated to gain insights into the origins of fractality. In this work, we introduce two novel network models, to better understand the growing mechanism and structural characteristics of fractal networks. The Repulsion Based Fractal Model (RBFM) is built on the well-known Song-Havlin-Makse (SHM) model, but in RBFM repulsion is always present among a specific group of nodes. The model resolves the contradiction between the SHM model and the Hub Attraction Dynamical Growth model, by showing that repulsion is the characteristic that induces fractality.
The Lattice Small-world Transition Model (LSwTM) was motivated by the fact that repulsion directly influences the node distances. Through LSwTM we study the fractal-small-world transition. The model illustrates the transition on a fixed number of nodes and edges using a preferential-attachment-based edge rewiring process. It shows that a small average distance works against fractal scaling, and also demonstrates that fractality is not a dichotomous property, continuous transition can be observed between the pure fractal and non-fractal characteristics.

\end{abstract}
\section{Introduction}
Modelling real networks has attracted a great deal of research in the last two decades since mathematical models allow us to rigorously and extensively investigate the underlying mechanisms of networks, discover substantial network properties, and shed light on their origins. For example, the preferential attachment model of Barabási and Albert explained the origin of the scale-free property~\cite{barabasi1999emergence}, the model of Watts and Strogatz helped in understanding the "small-world" phenomena in a variety of networks~\cite{watts1998collective}, while the model of Newman provided a better understanding of the properties of highly clustered networks~\cite{newman2003properties}.

Fractality is another well-studied property, which is present in a large number of real networks~\cite{rosenberg2020fractal,wen2021fractal}. Fractal scaling of networks has been introduced by Song \textit{et al.}~\cite{song2005self} motivated by the notion of geometric fractals. Fractality is defined by the so-called box-covering method. A network is called fractal, if the minimum number of boxes required to cover the whole vertex set follows a power-law relation with the size of the boxes, i.e.:
\[
N_B(l_B)\sim l_B^{-d_B},
\]
where $l_B$ denotes the size of the boxes, $N_B(l_B)$ stands for the number of $l_B$-sized boxes resulting from box-covering, and $d_B$ is called the box-dimension or fractal dimension of the network (if exists).

After laying the foundation of fractal network analysis, Song \textit{et al.} also proposed a mathematical model to explain the emergence of fractality in complex networks \cite{SHM_origins_sw9}. The main steps of the Song-Havlin-Makse (SHM) model can be summarised as follows:
\begin{enumerate}
    \item \label{shm:1} The initial graph is a simple structure, e.g., two nodes connected via a link.
    \item \label{shm:2} In iteration step $t+1$ we connect $m$ offspring to both endpoints of every edge, i.e., a $v$ node gains $m \cdot \deg_t(v)$ offspring, where $m$ is a predefined parameter and $\deg_t(v)$ is the degree of node $v$ at the end of step $t$.
    \item In iteration step $t+1$ every $(u,v)$ edge is removed independently with probability $p$, where $p$ is a predefined parameter. When an edge is removed, it is replaced with a new edge between random offspring of $u$ and $v$.
\end{enumerate}
The network grows dynamically and the degree correlation (hub repulsion/att-raction) of the emerging graph is driven by parameter $p$. The fractality is also influenced by the choice of parameter $p$, namely, it can be shown that the generated network is fractal for $p=1$, and non-fractal for $p=0$ \cite{SHM_origins_sw9}. The intermediate values develop mixtures between the two properties. The authors conclude that the \say{repulsion-between-hubs} principle is the key to the emergence of fractal scaling~\cite{SHM_origins_sw9}.

Kuang \textit{et al.} proposed the Hub Attraction Dynamical Growth (HADG) model, which is a modification of the SHM model, where the novelty lies in the flexible edge rewiring probability \cite{HADGM}. They demonstrated that by assigning smaller rewiring probability to edges connecting hubs, it is possible to create fractal networks with hub attraction behaviour. They also introduced a so-called \say{within-box link-growth} phase to the model to increase the clustering coefficient of the resulting network, which does not affect the fractal scaling \cite{HADGM}. 

Besides the relation of hub repulsion/disassortativity to fractality, another interesting phenomenon to model has been the conflicting relation of fractality and the small-world property \cite{kawasaki2010reciprocal}. For example, Rozenfeld \textit{et al.} proposed a new family of recursive networks, which are small-world for certain parameter settings, and fractal for others \cite{uvflower_sw8}. There are also many articles, which introduce models that exhibit a transition between the two properties \cite{li2017fractal_sw3,rozenfeld2010small_sw10,watanabe2015fractal_sw4,zhang2008transition_sw15}.

In this work, we introduce two novel fractal network models. First, we present the Repulsion Based Fractal Model (RBFM), which is intended to resolve the seeming contradiction of \cite{SHM_origins_sw9} and \cite{HADGM} by showing that repulsion causes the fractality of both the SHM and HADG models. Motivated by the fact that repulsion between nodes inevitably increases the average shortest path distance of a network, we introduce a second model to study the relationship between fractality and small-worldness. The second model, called Lattice Small-world Transition Model (LSwTM), supports the findings of earlier works \cite{li2017fractal_sw3,rozenfeld2010small_sw10,watanabe2015fractal_sw4} that small-world property interferes with fractality, and that real transition exists between the two characteristics. In contrast to the related works, our model is not relying on the normalisation method, and LSwTM is not a growing network, but the transition is shown on a fixed number of edges and nodes.

\section{Repulsion Based Fractal Model}
This model is based on the Song-Havlin-Makse model, it also evolves through time, and we rewire edges to create repulsion among nodes. However, here the probability of an edge to be rewired is not fixed but depends on the degree of its endpoints. In contrast to the Song-Havlin-Makse model, repulsion is always present in RBFM, moreover, with a predefined parameter we can specify the nodes that repel each other (e.g., hubs or small degree nodes). 
The model also adapts the \say{within-box link-growth} step of Kuang \textit{et al.}~\cite{HADGM} in order to create more realistic networks.
The growing mechanism of the Repulsion Based Fractal Model is as follows:
\begin{enumerate}
    \item Similarly to the Song-Havlin-Makse model, we start with a simple graph structure, e.g. two nodes connected via a link.
    \item The growth process of the model is the same as step \ref{shm:2} of the SHM model, namely in iteration step $t+1$ we connect $m \cdot \deg_t(v)$ offspring to every $v$ node, where $m$ is a predefined parameter and $\deg_t(v)$ is the degree of node $v$ at the end of step $t$.
    \item In iteration step $t+1$ we remove every $(u,v)$ edge with probability $p_{uv}^Y$ that depends on the mean degree of $u$ and $v$ normalised by the maximum degree:
    \[
    p_{uv}^Y = 1 - \left|Y - \frac{\deg_t(u)+\deg_t(v)}{2\cdot \deg_{t, \max}}\right|,
    \]
    where $Y \in [0,1]$ is a predefined parameter, $\deg_t(u)$ is the degree of node $u$, $\deg_{t, \max}$ is the maximum degree at step $t$. When an edge is removed, it is replaced with a uniformly randomly chosen new edge between the offspring of its endpoints.
    \item \label{RBFM:3} We add $\deg_t(v)$ edges among the newly generated offspring of every old node $v$. In order not to create self-loops this step is only executed, when $m>1$.
\end{enumerate}

\begin{figure}[]
    \centering
    \includegraphics[width=\textwidth]{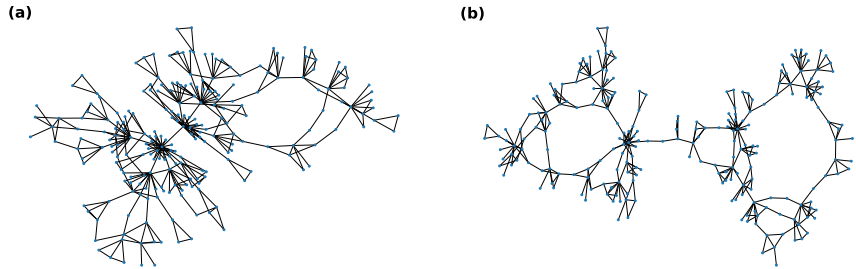}
    \caption{Illustration of the Repulsion Based Fractal Model for \textbf{(a)} $Y=0$, i.e., when small degree nodes tend to repel each other, \textbf{(b)} $Y=1$, i.e., when the repulsion is present among hubs.}
    \label{fig:illustration_rbfm}
\end{figure}

With the Y parameter, we assign high edge rewiring probability to those edges, which endpoints’ average degree is close to $Y \cdot \deg_{t, \max}$. For example, if $Y = 0$, with high probability we rewire those edges, which connect nodes with a relatively small degree, on the other hand in the case of $Y = 1$, with high probability we rewire the edges, that are linked between nodes with large degree (hubs). Figure~\ref{fig:illustration_rbfm} illustrates these two extreme cases of the model. The speciality of this model, is that it gives rise to fractal graphs for all $Y \in [0, 1]$, as it can be seen on Figure \ref{fig:rbfm_contour}\textbf{(a)}, too.

According to Song \textit{et al.}~\cite{SHM_origins_sw9} fractality is driven by disassortativity (negative degree correlation), however, Kuang \textit{et al.}~\cite{HADGM} introduced a model that generates fractal networks where the hubs are connected. The Repulsion Based Fractal (RBF) Model suggests that the property, which affects the fractal scaling of a network is \say{repulsion}, and repulsion does not necessarily have to be among hubs. The resolution of the contradiction lies in the fact that repulsion clearly affects the correlation of degrees. If there is a repulsion between hubs, i.e., if in the RBF model $Y = 1$, then hubs are only connected with small degree nodes, thus the degrees are anti-correlated and the network is disassortative. On the other hand, when the repulsion is between the small degree nodes, there are long paths consisting of small degree nodes, but in this case, the hubs are connected, hence there is a significantly larger correlation between the degrees. However, the resulting network is still fractal.
 The common mechanism that drives the fractality of the Song-Havlin-Makse model \cite{SHM_origins_sw9}, the model of Kuang \textit{et al.} \cite{HADGM}, and the RBFM is repulsion.

Clearly, repulsion makes the graphs spread out, i.e., when the nodes are repelling each other, then the graph cannot be too compact. To study the relationship between small-worldness and fractality we introduce the Lattice Small-world Transition model that is detailed in Section \ref{lswtm}.

\subsection{Properties of the RBFM}
Some of the main properties of the networks generated by the Repulsion Based Fractal Model are deterministic, i.e., do not depend on the exact realisations, but are determined by the parameters. In fact, the number of nodes and edges of the network are only influenced by the choice of parameter $m$ and the number of iterations $t$. Following the notations and thread of \cite{HADGM,SHM_origins_sw9} we can conclude the following for the $m>1$ case of the model:
\begin{align*}
    E(t)-E(t-1)&=2m\cdot E(t-1) + 2\cdot E(t-1)\\
    E(t)&=(2m+3)\cdot E(t-1)
    = E(0)\cdot (2m+3)^t,
\end{align*}
where $E(t)$ denotes the number of edges of the network at step $t$, while $E(0)$ is the number of edges of the initial graph. For the number of nodes the following findings can be made:
\begin{align*}
    N(t)-N(t-1)&=2m\cdot E(t-1)\\
    N(t)&=N(t-1)+2m\cdot E(0)\cdot (2m+3)^{t-1}\\
    &=N(0)+E(0)\cdot 2m \cdot \sum_{k=0}^{t-1}{(2m+3)^k}\\
    &=N(0)+E(0)\cdot \frac{m}{m+1} \cdot ((2m+3)^t-1),
\end{align*}
where $N(t)$ refers to the number of nodes of the network at iteration step $t$, and $N(0)$ is the number of nodes of the initial graph.

When $m=1$, step \ref{RBFM:3} cannot be executed, consequently the previous derivations simplify:
\begin{align*}
    E(t)-E(t-1)&=2m\cdot E(t-1)\\
    E(t)&= E(0)\cdot (2m+1)^t = E(0)\cdot 3^t \\
    N(t)&=N(0)+E(0)\cdot 2m \cdot \sum_{k=0}^{t-1}{(2m+1)^k} \\
    &= N(0)+ E(0)\cdot ((2m+1)^t-1)= N(0)+ E(0)\cdot (3^t-1)
\end{align*}

Since the number of nodes and edges are deterministic in parameters $m$ and $t$, the average degree of the network can also be studied analytically. In the $m=1$ case:
\begin{align*}
    d_{avg} = \frac{2\cdot E(0) \cdot 3^t}{N(0)+ E(0)\cdot (3^t-1)} \xrightarrow[t \to \infty]{} 2
\end{align*}
When $m>1$:
\begin{align*}
    d_{avg} = \frac{2\cdot E(0) \cdot (2m+3)^t}{N(0)+E(0)\cdot \frac{m}{m+1} \cdot ((2m+3)^t-1)} \xrightarrow[t \to \infty]{} \frac{2(m+1)}{m} \xrightarrow[m \to \infty]{} 2
\end{align*}

To investigate how similar the random generalisations of the model with a given parameter setting are in terms of various network characteristics, we generated 30 graphs with a given parameter setting.
We examined 7 network metrics, namely, the average path length, the normalised diameter, the normalised maximum degree, the average clustering coefficient, the assortativity coefficient, the maximum of the eigenvector centralities, and the skewness of the degree distribution. The results can be found in the supplementary material: \url{https://github.com/marcessz/fractal-network-models}. The examined characteristics can be considered stable because the values of the metrics do not differ significantly for the different realisations of the networks. The average clustering coefficient, and also the maximum of the eigenvector centralities may not seem to be as consistent as the other metrics, but the range, in which the values vary is still quite small. Furthermore, the fluctuation decreases for larger networks, i.e., when the model performs more iterations. Overall, we can conclude that the main characteristics of the network model for a given parameter setting do not depend heavily on the exact realisations.

\begin{figure}[]
    \centering
    \includegraphics[width=0.49\textwidth]{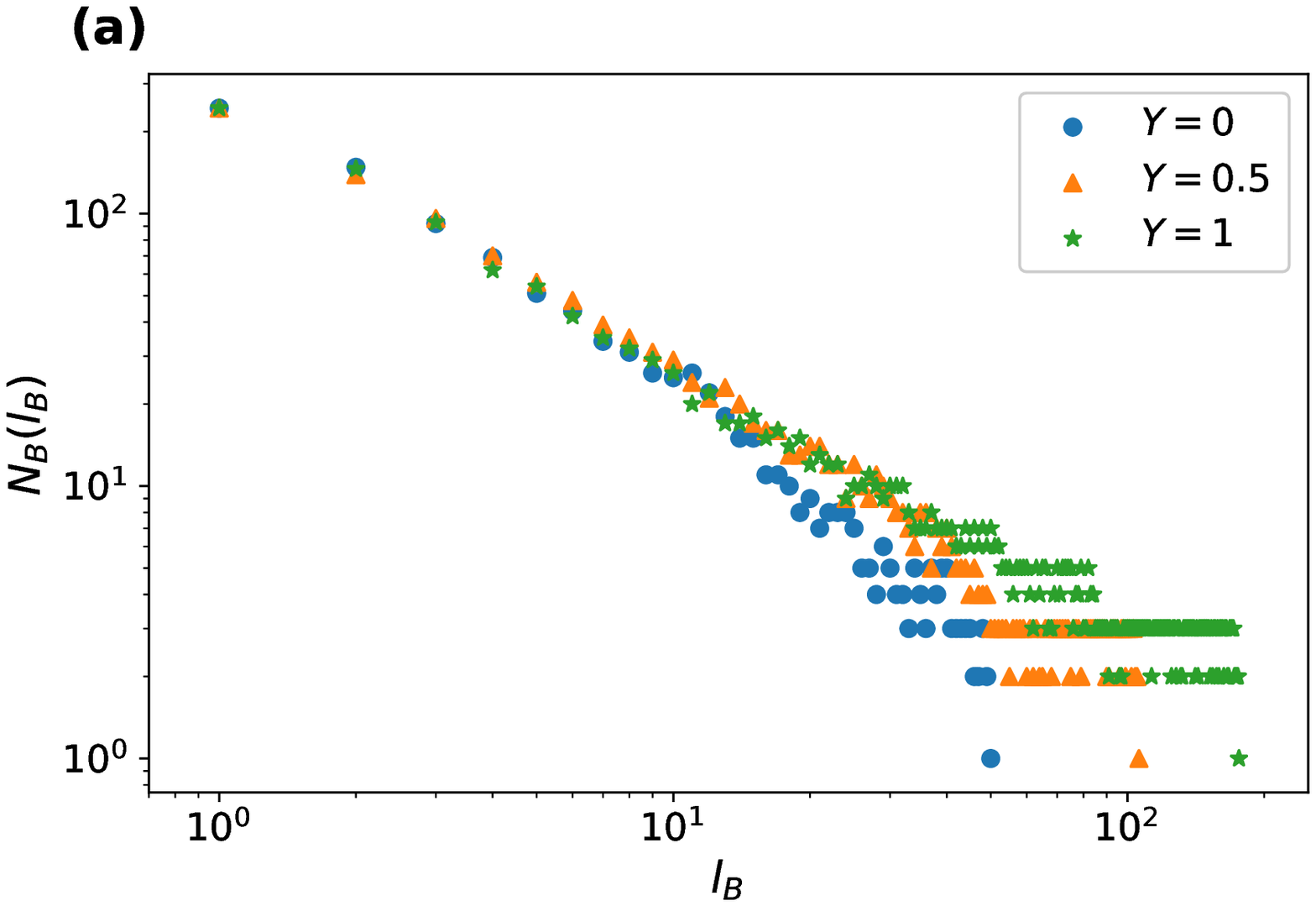}
    \includegraphics[width=0.49\textwidth]{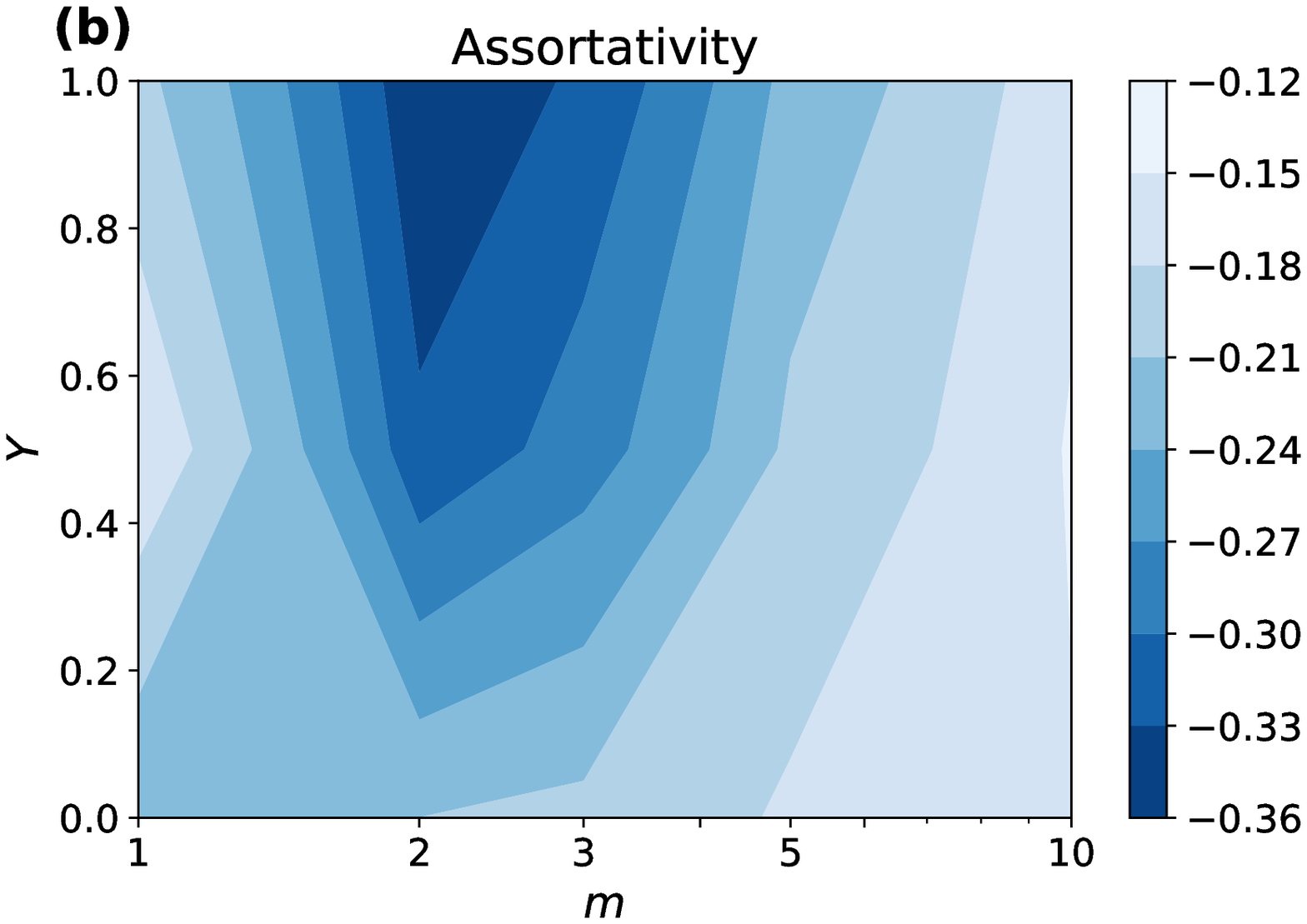}\\
    \includegraphics[width=0.49\textwidth]{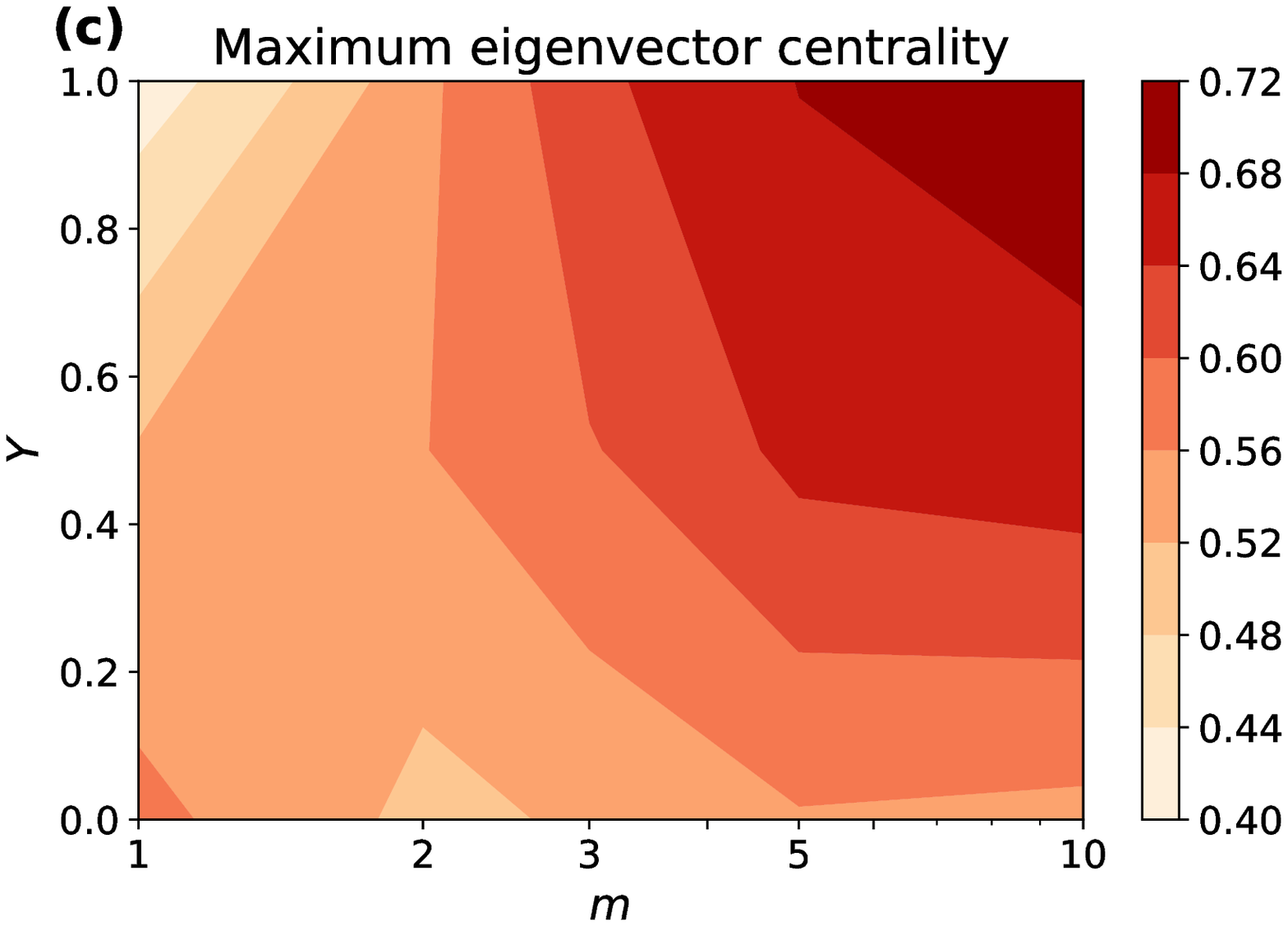}
    \includegraphics[width=0.49\textwidth]{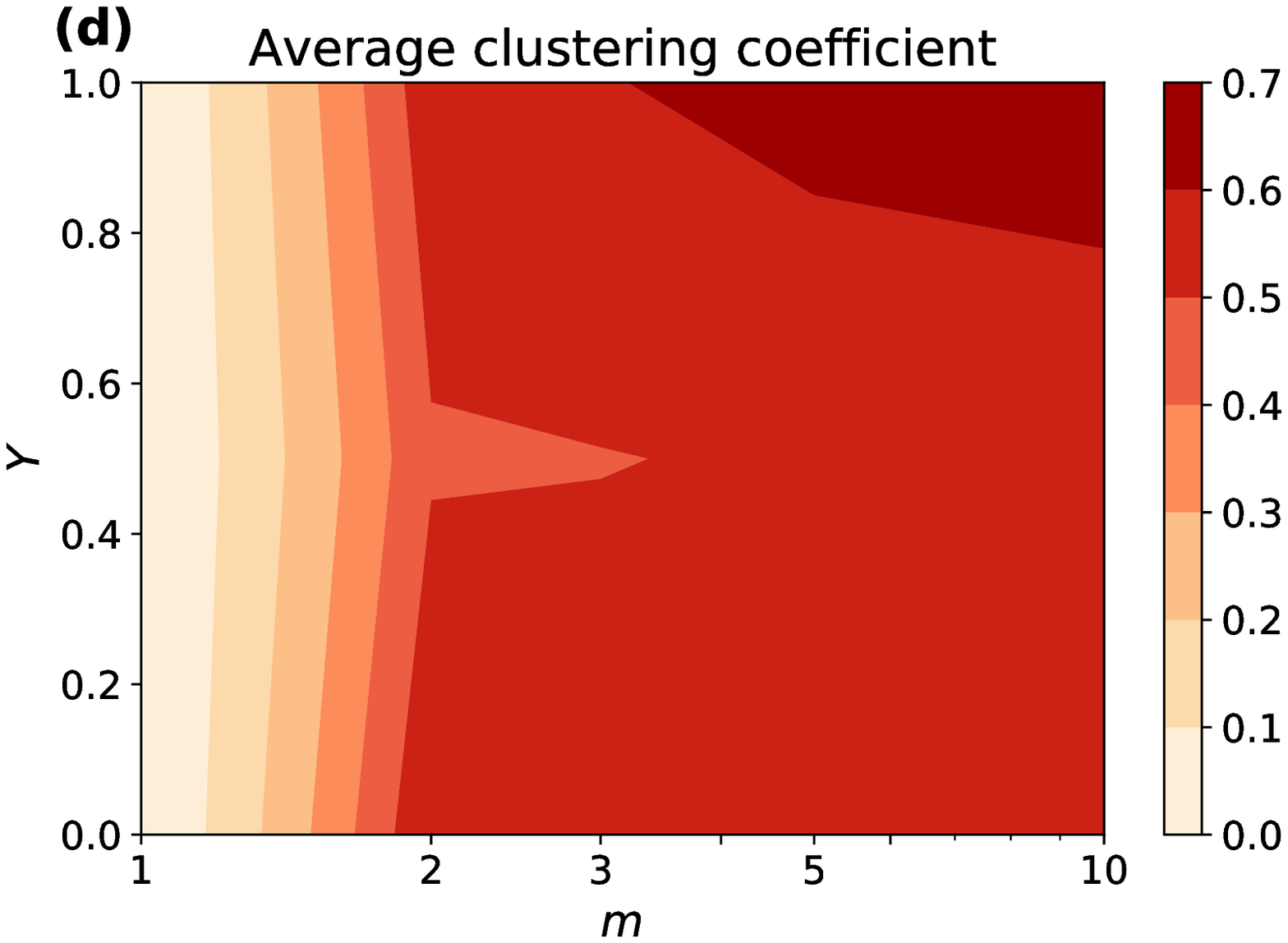}
    \caption{\textbf{(a)} Illustration of the fractality of the Repulsion Based Fractal Model for different parameter settings. 
    \textbf{(b)}-\textbf{(d)} Contour plots of three network metrics, which show the change of the metric values as a function of parameters $m$ and $Y$ of the Repulsion Based Fractal Model. The subfigures illustrate the values of the \textbf{(b)} assortativity coefficient, \textbf{(c)} maximal eigenvector centrality, \textbf{(d)} average clustering coefficient for networks generated by the model with $t=3$ setting and with multiple choices of parameters $m$ and $Y$.}
    \label{fig:rbfm_contour}
\end{figure}

We also investigated how sensitive the model is to its parameters, i.e., how a small change in the parameters affects the characteristics of the network. Due to the complicated network evolution process, it is difficult to analytically determine various characteristics of the model, thus we generated 30 graphs with a certain parameter setting and averaged the graph metrics of these 30 realisations. We repeated this procedure for various parameter settings to assess the parameter sensitivity of the model.  Figure \ref{fig:rbfm_contour} \textbf{(b)}, \textbf{(c)}, and \textbf{(d)} show the contour plots of three network metrics (assortativity, maximum eigenvector centrality, average clustering coefficient). It can be seen that some of the characteristics do not, or just barely depend on the choice of parameter $Y$, while others are highly influenced by it. The average clustering coefficient is quite non-sensitive for all of the parameters, but naturally increases greatly, when the structure of the network is more complex than a path (i.e., when $m>1$). The other examined characteristics highly depend on the realisation of parameter $Y$. The average path length and the (normalised) diameter become larger as we increase the value of $Y$, and similar holds for the maximal eigenvector centrality. The generated networks are disassortative in all of the cases, but the model gets more disassortative for larger values of $Y$, i.e. when the repulsion is created among large degree nodes.

\section{Lattice Small-world Transition Model}\label{lswtm}
Our novel model embraces both preferential attachment mechanism and the \say{geometric} structure that emerges in networks that can be embedded in two- and three-dimensional Euclidean spaces, for instance, infrastructure networks \cite{csanyi2004fractal_sw12}, blood vessels, and trabecular bones \cite{wlczek1992fractal}. Several network models have been introduced to create a synergy between geometric network models and preferential attachment. For example, Flaxman \textit{et al.} have presented two growth models in which the vertices of the network are randomly chosen points of the three-dimensional unit sphere, and edges are created taking into account both the proximity of the nodes and an extended preferential attachment mechanism \cite{flaxman2006geometric,flaxman2007geometric}. The use of the preferential attachment principle to create small-world networks has also been studied extensively \cite{jian2006multistage,wang2008evolving}.

Here, we present a network model, which utilises the fractal nature of grid-like structures, and at the same time works against it with the preferential attachment mechanism. The Lattice Small-world Transition Model is defined as follows:
\begin{enumerate}
    \item We start with a $d$-dimensional (practically $d=2$) grid graph with $n_1 \times n_2 \times \dots \times n_k$ vertices.
    \item With probability $p$, every $(v_i, v_j)$ edge is replaced by $(v_i, v_k)$, where $v_k$ is chosen with a probability, that is proportional to $p_{v_k}$:
    \[
    p_{v_k} = \frac{1}{1 + \exp \left( -a \cdot \left(\frac{\deg(v_k)}{\deg_{\max}} - \frac{1}{2}\right)\right) }, 
    \]
    where $a$ is a positive constant, $\deg(v_k)$ is the degree of node $v_k$ and $\deg_{\max}$ is the maximum degree of the current graph. Note that when the normalised degree of $v_k$ is $\frac{1}{2}$, then $p_{v_k}$ equals $\frac{1}{2}$. If $\deg(v_k)$ is less than $\deg_{\max}/2$, then $p_{v_k}$ is close to zero, on the other hand, when $\deg(v_k) > \deg_{\max}/2$ then $p_{v_k}$ is nearly one (if $a$ is large enough).
    By practical motivation, to avoid multiple edges, the set of nodes to select $v_k$ from is defined as $S_{v_i}~=~V\backslash\{\Gamma_{v_i}\cup \{v_i\}\}$, where $\Gamma_{v_i}$ denotes the neighbourhood of $v_i$.
    By default, $v_j$ is replaced with $v_k$ during the rewiring process, however, if in this way the graph becomes disconnected, $v_i$ is replaced instead.
\end{enumerate}

\begin{figure}[]
    \centering
    \includegraphics[width=\textwidth]{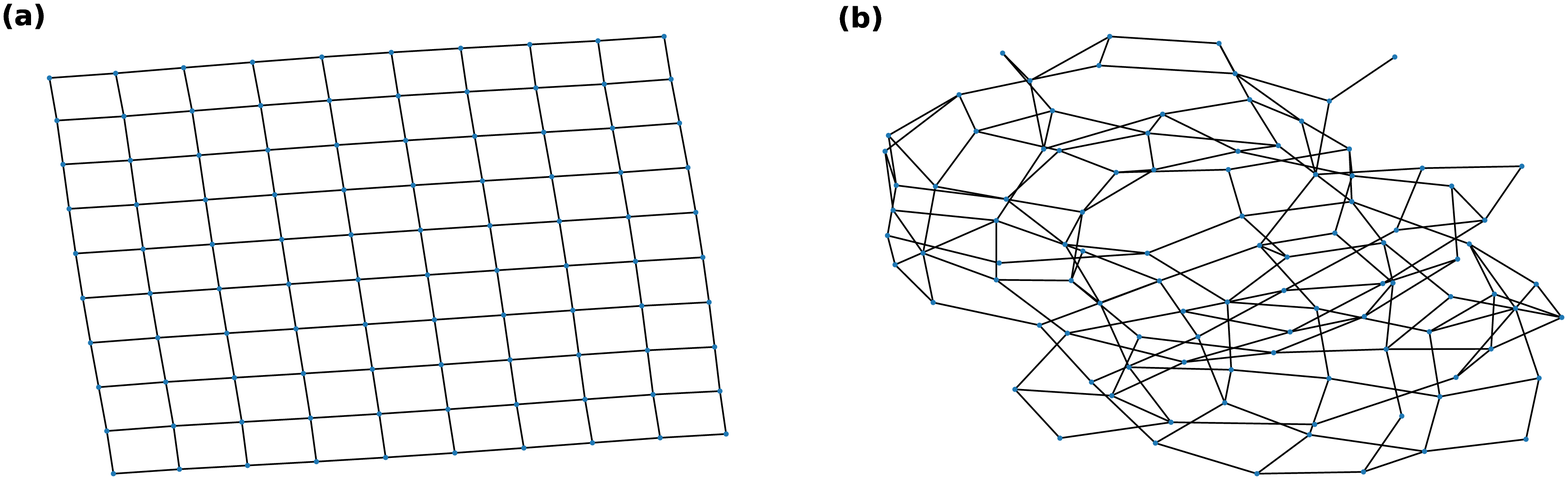}
    \caption{Illustration of the Lattice Small-world Transition Model for \textbf{(a)} $p=0$, \textbf{(b)} $p=0.1$.}
    \label{fig:illustration_mixture}
\end{figure}

\begin{figure}[]
    \centering
    \includegraphics[width=\textwidth]{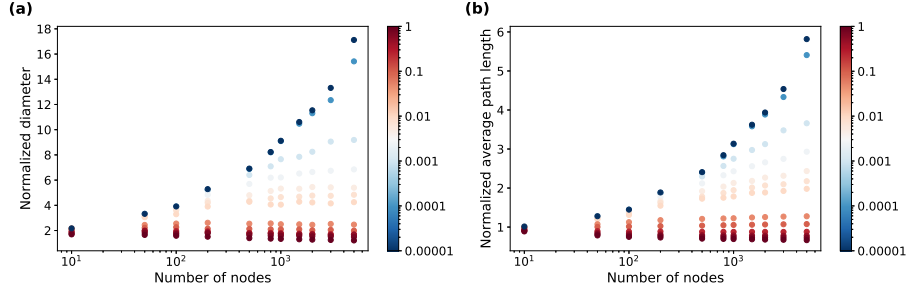}
    \caption{(\textbf{a}) Normalised diameter and \textbf{(b)} average path length (i.e. diameter/average path length divided by the logarithm of the size) as a function of the logarithm of the number of nodes.  The colouring is based on the $p$ parameter of the model.}
    \label{fig:mixture_diam_avgpath}
\end{figure}

Figure \ref{fig:illustration_mixture} illustrates that even a small rewiring probability results in a network that differs greatly from a grid graph. The fractality of the generated network depends on the choice of $p$. For $p=0$ the network is purely fractal, and as $p$ grows the model shows a transition from fractal to non-fractal. It also has to be mentioned that this transition is not sharp, and there are intermediate states, where the network is locally fractal, although the pure property is no longer present globally. These properties are well illustrated on Figure \ref{fig:mixture_contour}\textbf{(a)}. Furthermore, as fractality disappears small-world property arises. Figure \ref{fig:mixture_diam_avgpath} shows the change in the normalised diameter and average path length in terms of the model parameters. The normalisation by the logarithm of the number of nodes is done to be able to compare the distances of networks of different sizes. It can be seen that the distances are growing as $p$ decreases, i.e., as the networks become fractal. Both RBFM and LSwTM suggest that a fractal network has to be spread out, and as LSwTM illustrates, as we rewire edges according to the preferential attachment mechanism, it turns small-world and loses its fractal structure.

\subsection{Properties of the LSwTM}
Similarly to the RBFM, some properties of the networks generated by the Lattice Small-world Transition Model are deterministic in the model parameters. The number of nodes and edges of a grid graph is determined by its $n_i \; (i=1, 2, \ldots, d)$ parameters. Since the model only includes edge rewiring steps and there is no edge/node deletion or addition, the resulting network has the same number of nodes and edges as the initial graph.

Simple derivations can be made for the number of nodes and edges of the generated networks, when the initial grid graph is $d$-dimensional, i.e when we have parameters $n_1, n_2, \ldots, n_d$:
\begin{align*}
    |V| &= \prod_{i=1}^{d}{n_i} \\
    |E| &= \sum_{i=1}^{d}{(n_i-1)\cdot \frac{\prod_{j=1}^{d}{n_j}}{n_i}}
    \end{align*}
 \begin{align*}  
    d_{avg} &= \frac{2\cdot \sum_{i=1}^{d}{(n_i-1)\cdot \frac{\prod_{j=1}^{d}{n_j}}{n_i}}}{\prod_{i=1}^{d}{n_i}}
    = \frac{2\cdot \left( \sum_{i=1}^{d}{ \prod_{j=1}^{d}{n_j}} - \sum_{i=1}^{d}{\frac{\prod_{j=1}^{d}{n_j}}{n_i}}\right)}{\prod_{i=1}^{d}{n_i}}\\
    &= 2d - 2\cdot \sum_{i=1}^{d}{\frac{1}{n_i}}
\end{align*}
When all $n_i$s are large, the average degree is close to $2d$. On the other hand, when $n_i=2$ for all $i=1, 2, \ldots, d$, the average degree equals to $d$. Consequently, if $n_i>1$ for all $i$, the following bounds hold: $
    d \leq d_{avg} < 2d. $
If $n_i=1$ for at least one $i$, the average degree can be smaller than $d$, since in this case we basically start with a lattice of dimension less than $d$.

We also investigated how similar the random realisations of the LSwTM are, moreover we also executed a sensitivity analysis on the parameters of the model. Again, we consider the same seven structural network metrics as before. For a given parameter setting the generated networks have very similar properties concerning the examined characteristics. The results can be found in the supplementary material: \url{https://github.com/marcessz/fractal-network-models}.

\begin{figure}[]
    \centering
    \includegraphics[width=0.49\textwidth]{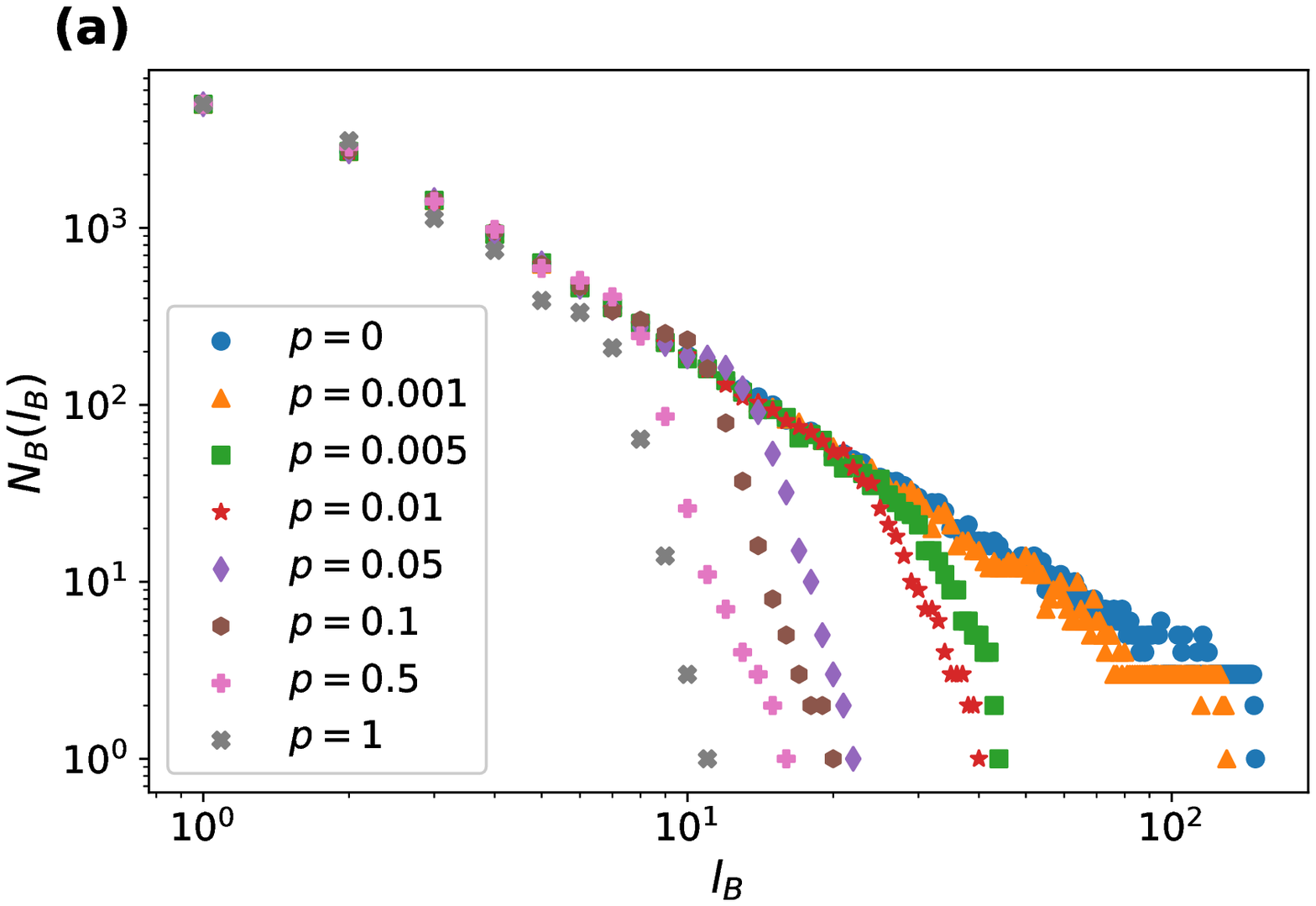}
    \includegraphics[width=0.49\textwidth]{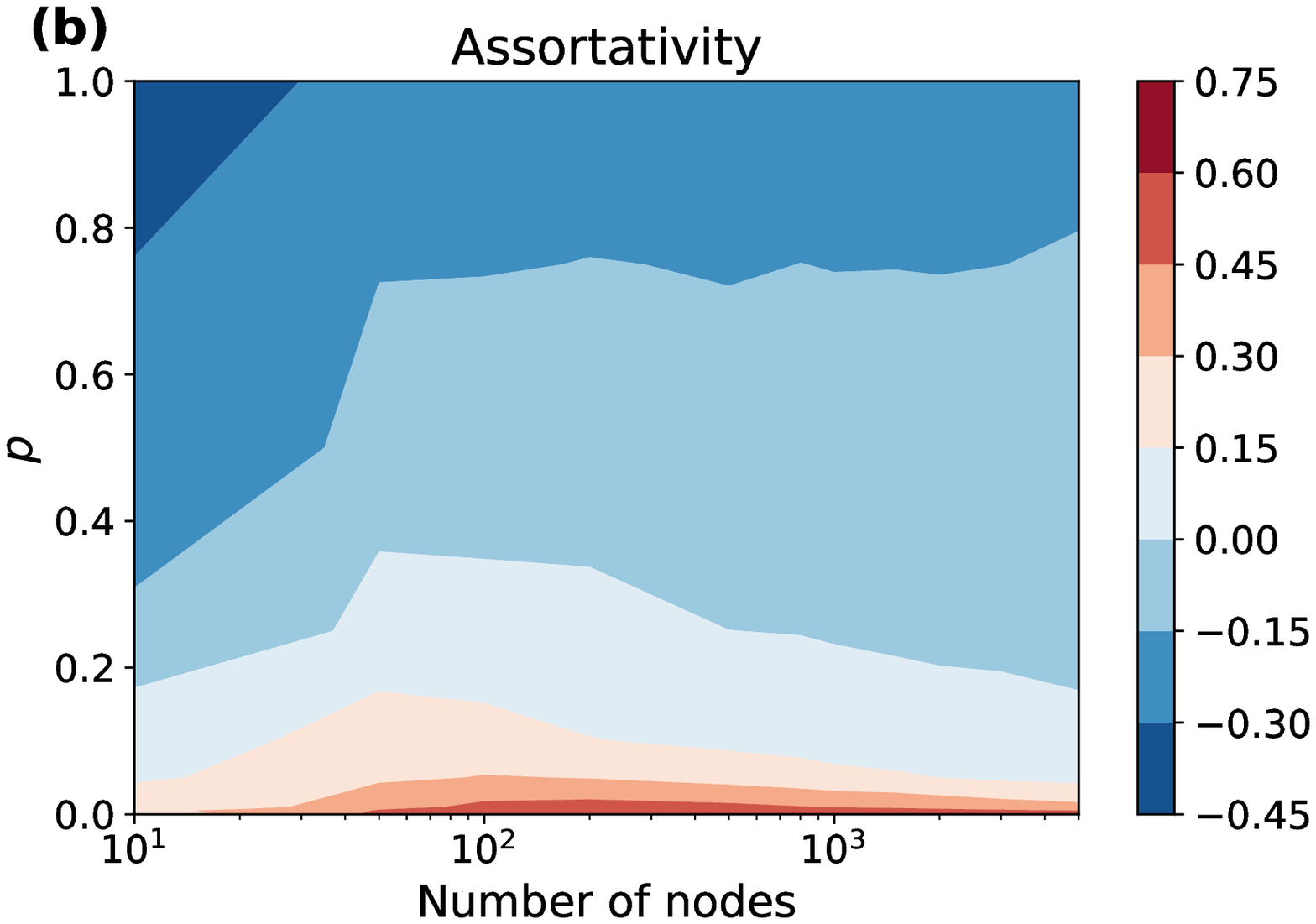}\\
    \includegraphics[width=0.49\textwidth]{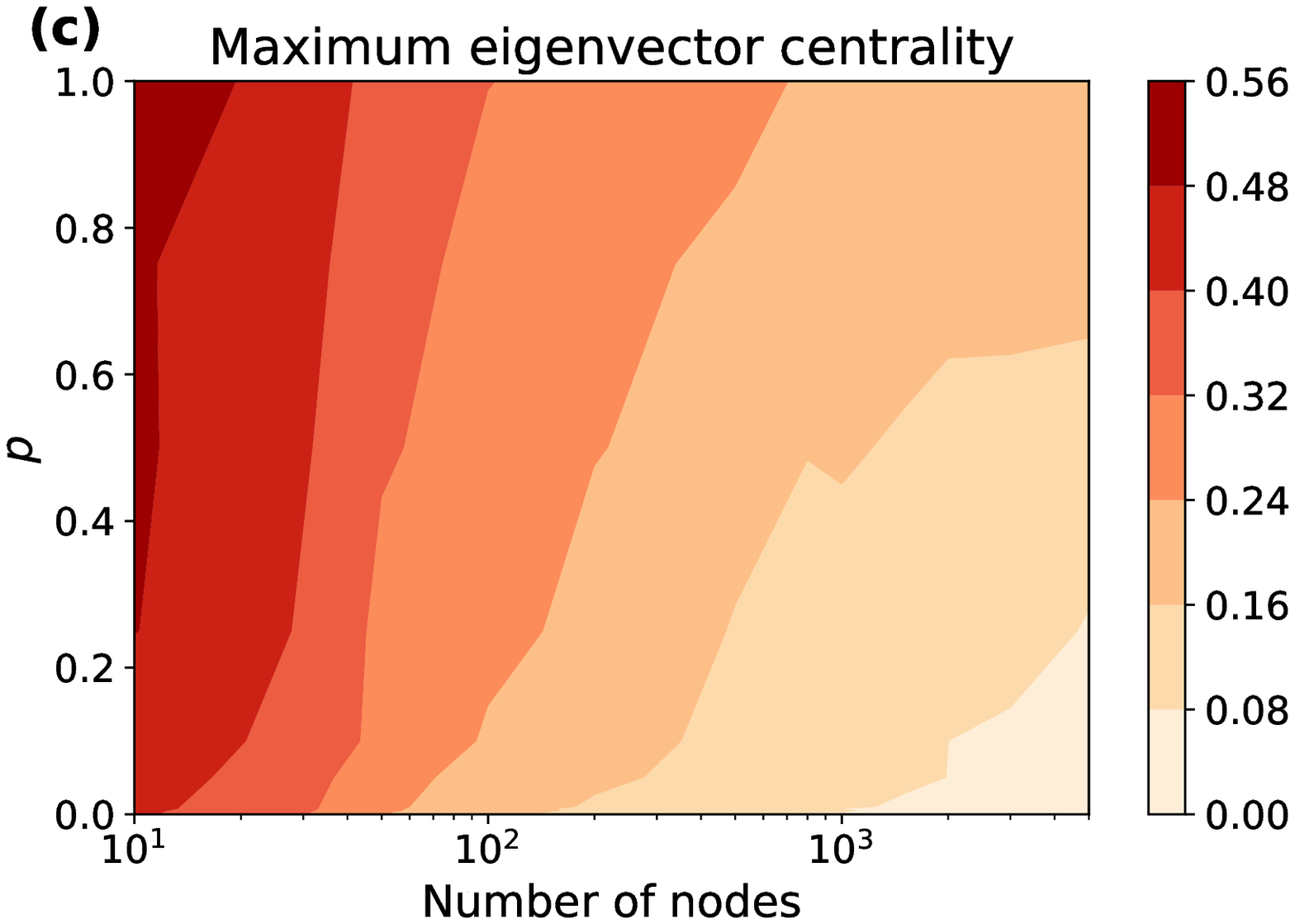}
    \includegraphics[width=0.49\textwidth]{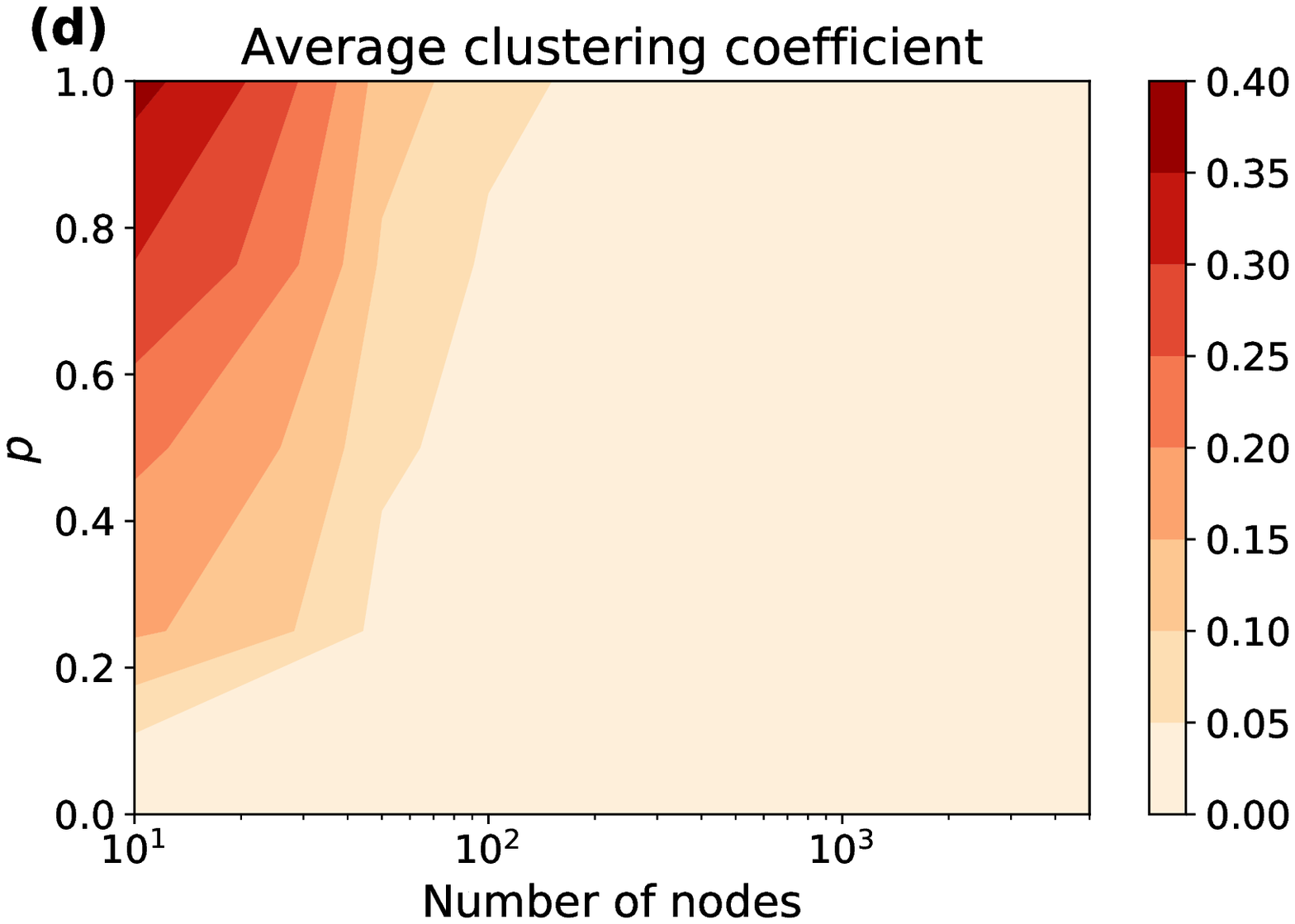}
    \caption{\textbf{(a)} Illustration of the fractality of the LSwT model for different parameter settings. \textbf{(b)}-\textbf{(d)} Contour plots of three graph metrics, illustrating the metric values as a function of the parameters of the LSwTM. The subfigures illustrate the values of the \textbf{(b)} assortativity coefficient, \textbf{(c)} maximal eigenvector centrality, \textbf{(d)} average clustering coefficient for networks generated by the model with multiple choices of the number of the nodes and parameter $p$.}
    \label{fig:mixture_contour}
\end{figure}

Figure~\ref{fig:mixture_contour} \textbf{(b)}, \textbf{(c)}, and \textbf{(d)} show how the model parameters affect some characteristics of the network. Most of the graph metrics are influenced mainly by the network size, for example, the maximal eigenvector centrality decreases as the network grows. The average clustering coefficient, apart from the small networks, is around 0, independently of the value of $p$. Some characteristics, however, rather depend on parameter $p$ and are not affected highly by the network size. The assortativity coefficient decreases with the growing values of $p$, and the same holds for the average path length and the (normalised) diameter too, with the remark that these two become small even for values of $p$ slightly greater than 0.

\section{Discussion and Summary}
In this work, we introduced and analysed two network models to better understand what mechanisms affect the fractality of networks.

The Repulsion Based Fractal Model is based on the models of Song \textit{et al.}~\cite{SHM_origins_sw9} and Kuang \textit{et al.}~\cite{HADGM}. Song \textit{et al.} assumed that in fractal networks the hubs are not connected, in other words, there is a repulsion between hubs, which is also known as disassortative mixing. On the other hand, Kuang \textit{et al.} modified the Song-Havlin-Makse model, in such a way that it is able to generate fractal networks with connected hubs. Although Kuang \textit{et al.} pointed out that disassortativity is not the mechanism that makes a network fractal, they did not investigate thoroughly the origins of fractality. Through the Repulsion Based Fractal Model, we showed that the repulsion between nodes induces fractality, and the repelling nodes do not necessarily have to be hubs. The RBF model also well illustrates that repulsion affects not only the fractality but the assortative-mixing of a network, which resolves the contradiction between the findings of Song \textit{et al.}~\cite{SHM_origins_sw9} and Kuang \textit{et al.}~\cite{HADGM}.

As the RBFM and earlier works suggest \cite{li2017fractal_sw3,rozenfeld2010small_sw10,watanabe2015fractal_sw4,zhang2008transition_sw15}, fractality is influenced by the node distances. We introduced a model, which shows that if we take a purely fractal network, and we start to rewire edges according to the preferential attachment principle, then as the shortest path length decreases, the fractal structure breaks down gradually, and eventually the small-world property will dominate the network.
%
%

%
%
%
\bibliographystyle{splncs04}
\bibliography{bibliography}

\end{document}